%% file: paper.tex
\documentclass[useAMS,usenatbib]{mn2e}
\usepackage{epsf}
\usepackage{amsmath,amssymb}

\newcommand{\Ms}{M_\star} \newcommand{\Rs}{R_\star}

\newcommand{\Mbh}{M_{\bullet}} \newcommand{\Mo}{M_{\odot}}

\newcommand{\Ro}{R_{\odot}}

\def\apgt{\ {\raise-.5ex\hbox{$\buildrel>\over\sim$}}\ } 
\def\aplt{\ {\raise-.5ex\hbox{$\buildrel<\over\sim$}}\ }

\bibpunct{\hspace{-4pt}}{}{}{a}{}{}

\title[Tidal Capture of Stars by Intermediate-Mass Black Holes]
{
  Tidal Capture of Stars by Intermediate-Mass Black Holes
}

\author[H. Baumgardt, C. Hopman and S. Portegies Zwart, J. Makino]
{
H. Baumgardt$^{1}$\thanks{e-mail: holger@astro.uni-bonn.de (HB);
   clovis.hopman@weizmann.ac.il (CH); spz@science.uva.nl (SPZ);
   makino@astron.s.u-tokyo.ac.jp (JM)}, 
C. Hopman$^{2}$\footnotemark[1],
S. Portegies Zwart$^{3, 4}$\footnotemark[1] and
J. Makino$^5$\footnotemark[1]\\ 
$^{1}$Sternwarte, University of Bonn, Auf dem H\"ugel 71, 53121 Bonn,
      Germany\\
$^{2}$Faculty of Physics, Weizmann Institute
      of Science, P.O. Box 26, Rehovot 76100, Israel\\
$^{3}$Astronomical Institute ``Anton Pannekoek,'', University of
      Amsterdam, Kruislaan 403, Netherlands\\
$^{4}$Section Computational Science, University of Amsterdam,
      Kruislaan 403, Netherlands\\
$^{5}$Department of Astronomy, University of Tokyo, 7-3-1 Hongo,
      Bunkyo-ku,Tokyo 113-0033, Japan
}

\begin{document}

\date{Accepted ????. Received ?????; in original form ?????}

\pagerange{\pageref{firstpage}--\pageref{lastpage}} \pubyear{2004}

\maketitle

\label{firstpage}

\begin{abstract}
Recent X-ray observations and theoretical modelling have made it
plausible that some ultraluminous X-ray sources (ULX) are powered by
intermediate-mass black holes (IMBHs).  N-body simulations have also
shown that runaway merging of stars in dense star clusters is a way to
form IMBHs. In the present paper we have performed $N$-body
simulations of young clusters such as MGG-11 of M82 in which
IMBHs form through runaway merging. We took into account
the effect of tidal heating of stars by the IMBH to study the tidal capture and disruption
of stars by IMBHs. Our results show that the IMBHs have a high chance
of capturing stars through tidal heating within a few core relaxation
times and we find that 1/3 of all runs contain a ULX within the age
limits of MGG-11, a result consistent with the fact that a ULX is
found in this galaxy. Our results strengthen the case for some ULX
being powered by intermediate-mass black holes.
\end{abstract}

\begin{keywords}
globular clusters: general -- black hole physics -- stellar dynamics
\end{keywords}

\section{Introduction}\label{sec:intro}
Ultra-luminous X-ray sources (ULX) are point-like X-ray sources
with isotropic X-ray luminosities in excess of $L=10^{40}$ erg/sec.
Various theories have been proposed in the literature concerning the
nature of ULX, like stellar-mass black hole binaries with mild
geometrical beaming (King et al.\ \cite{KI01}, Rappaport,
Podsiadlowski \& Pfahl \cite{RA05}) or high-redshift quasars
(Guti\'errez \& L\'opez-Corredoira \cite{GU05}) 
and it is likely that ULX are not a homogeneous class of objects
(Soria, Cropper \& Motch \cite{SO04}). 
Although luminous X-ray sources have been
observed in different kinds of environments, sources with luminosities of $L
\approx 2 \cdot 10^{39}$ erg/sec or brighter seem to be associated with young
stellar populations of star-forming galaxies (Irwin, Bregman \& Athey
\cite{IW04}; Swartz et al. \cite{S04}).

Since the Eddington luminosity
of a star of mass $M$ is given by
\begin{equation}
L_{\rm Edd}=1.3\times10^{38}{\rm erg s^{-1}}{M\over\Mo}
\end{equation}
where $M$ is the mass of the accreting object, most low-luminosity ULX
are probably stellar-mass black holes. However, there is mounting
evidence that the brightest ULX with luminosities exceeding
$L>10^{40}$ erg s$^{-1}$ could be intermediate-mass black holes
(IMBHs, see Miller \& Colbert \cite{MC04} for a review). The starburst
galaxy M82 for example hosts a ULX with brightness in the range
$L=(0.5-1.6) \cdot 10^{41} {\rm erg s^{-1}}$ (Matsumoto et al.\
\cite{MA01}, Kaaret et al.\ \cite{KA01}), corresponding to a black
hole with mass $350 -1200 M_\odot$ if emitting photons at the
Eddington luminosity.  The case for an IMBH in M82 is supported by a
54mHz quasi-periodic oscillation found in the X-ray flux (Strohmayer
\& Mushotzky \cite{SM03}) and also by the soft X-ray spectrum of this
source (Fiorito \& Titarchuk \cite{FT04}). Another argument supporting
the connection between ULX and IMBHs are cool thermal emission
components which have been found in the X-ray spectra of some ULX and
which can be fitted well with accretion disc models of IMBHs (Miller,
Fabian \& Miller \cite{MFM04}).

Portegies Zwart et al.\ (\cite{PZea04}) have performed $N$-body
simulations of the dense star cluster MGG-11 whose position coincides
with the ULX in M82. They found that runaway merging of massive
main-sequence stars in the centre of this cluster leads to the formation 
of a massive star with a mass of several 1000 $M_\odot$ within a few Myrs if the
initial concentration of the cluster is larger than that of a King
$W_0 = 9.0$ model. The connection between an IMBH and the ULX was
strengthened by Hopman, Portegies Zwart \& Alexander (\cite{HPZA04})
who showed through analytic estimates that an IMBH with a mass of 1000
$M_\odot$ residing in the centre of MGG-11 has a high probability of
tidally capturing a passing main-sequence star or giant.  The orbit of
the captured star was found to circularise even when dynamical
perturbations by other cluster stars were taken into account and the
system entered a Roche-lobe overflow (RLOF) phase once the captured
star had sufficiently increased its radius through stellar evolution.

Portegies Zwart, Dewi \& Maccarone (\cite{PZDM04}) and Li (\cite{L04})
showed that IMBHs with mass-transferring companions near the end of
their main-sequence lifetimes are able to create X-ray luminosities in
excess of $L = 10^{40} {\rm erg s^{-1}}$ over several Myrs, giving
them a high chance to be observed as ULX. For sufficiently low-mass
companions, the RLOF mass transfer stage might start after the host
cluster has been dissolved by the tidal field of the parent galaxy,
offering a way to explain isolated ULX (Hopman, Portegies Zwart \&
Alexander \cite{HPZA04}).

In this paper we present the results of detailed $N$-body simulations
of the formation and further evolution of an IMBH in a cluster with
characteristics similar to MGG-11 in the starburst Galaxy M82 (see
Portegies Zwart et al.\, \cite{PZea04}).  Our simulations included
the effects of tidal heating and the emission of gravitational waves
of passing stars and we investigated the chances that MGG-11 produces a
ULX within the age limits determined by observations (7-12 Myrs).

This paper is organised as follows: In section 2 we review the theory
of tidal heating and section 3 gives analytic estimates of tidal inspiral.
Section 4 describes the implementation of tidal heating into our
simulations. Section 5 discusses our results for MGG-11 and section 6
summarises the paper and presents our conclusions.

\section{Tidal heating of stars near a massive black hole}\label{sec:tidal}

Stellar systems are usually well described as a system of point
particles interacting with Newtonian gravity. In a few cases, however,
when the stellar density is sufficiently high, this approximation is
no longer accurate and fails to describe some of the important
processes which occur in stellar clusters. General relativistic
effects for example become important when stars approach each other at
distances close to their Schwarzschild radius, which can be relevant
for compact stars.  Main sequence stars are much larger and close encounters
can lead to hydrodynamical interactions, with stellar collisions as
the extreme. This may lead to a runaway merger in young and dense
star clusters (Portegies Zwart et al. \cite{PZea99}; Portegies Zwart
\& McMillan \cite{PZMcM02}; Freitag, G\"urkan \& Rasio
\cite{FR05}; see also section 5).

Stars that pass each other at somewhat larger distances may still affect
each others orbits by energy dissipation in a tidal interaction.  Such
an interaction may initiate tidal modes on both stars.  The energy
invested to raise the tides in the stars is taken from their kinetic
energies (Press \& Teukolsky \cite{PT77}).  Two initially unbound
stars may so become bound, or a bound eccentric binary can become
bound more tightly.

Stars can also have a tidal interaction with a black hole
(BH). The length-scale
for tidal forces of an IMBH of mass $\Mbh$ on a star of mass $\Ms$ and
radius $\Rs$ to become important is given by the tidal radius
(see e.g.\, Kochanek \cite{KO92}):
\begin{equation}\label{eq:rt}
  r_t \simeq \left(\frac{\Mbh}{\Ms}\right)^{1/3}\Rs.
\end{equation}

We now recognise several distinct regimes: (1) The 
pericenter distance is much larger than the tidal radius ($r_p\gg r_t$),
in which case tidal effects are negligible. (2) 
The distance of closest approach is smaller than the tidal
radius $r_p < r_t$. In this case 
the entire star may be disrupted or part of its
envelope may be stripped off to form a temporary accretion
disk around the black hole.  The gas accreting from the star
or the disc onto the black hole then causes a flare.  The
duration of the flare depends on how violent the interaction
was and whether or not an accretion disc was formed: for tidal
disruptions around super-massive black holes the flare might
last at most a few years (Rees \cite{R88}).
(3) In the intermediate regime ($r_p\gtrsim r_t$),
the star is tidally deformed, and orbital kinetic energy is
transferred to the internal energy of the star.  We call this the tidal
capture regime. 

For highly eccentric orbits ($e\gtrsim0.9$), the energy dissipation
per orbit can be parametrised by
\begin{equation}\label{eq:DEt}
   \Delta E_t = \frac{G\Ms^2}{\Rs } 
                \left( \frac{M_\bullet}{\Ms}
                \right)^2 
		\sum_{l=2}^\infty \left( \frac{\Rs}{r_p} 
		                  \right)^{2l+2} 
	        T_l(\eta),
\end{equation}
where
\begin{equation}
   \eta = \left( \frac{\Ms}{\Ms+M_\bullet} 
          \right)^{1/2} 
	  \left( \frac{r_p}{\Rs} 
	  \right)^{3/2} \;\; .
\end{equation}
Here $T_l(\eta)$ is the dimensionless tidal coupling function, which
depends on the stellar structure and is a strongly decreasing function
of the pericenter $r_p$ (see e.g. Press \& Teukolsky \cite{PT77}).  The
orbital energy at the tidal radius usually exceeds the binding energy
of the star by several orders of magnitude, and some fraction of the
orbital energy is dissipated at every new pericenter passage.
As a result the star becomes
very hot and expands. Both effects make it more luminous. The
stellar luminosity can become much larger than that achieved by
nuclear burning, and is of order
\begin{equation}
\label{eq:xiL}
   L_t = {\Delta E_t\over P}.
\end{equation}
Here $P$ is the orbital period of the binary system. 

It is unclear where a star stores the excess tidal energy. In the two
most extreme cases, the star can store the excess energy in the
surface layers, or in the bulk of the stellar material.  The first
case leads to high surface temperatures without appreciable expansion
of the star (McMillan, Dermott \& Taam \cite{McM87}), whereas the
latter causes the star to remain cold but expand dramatically
(Podsiadlowski \cite{P96}).  Alexander \& Morris \cite{AM03} dubbed
the two possibilities as ``hot'' and ``cold'' squeezars and argued
that these tidally excited stars may be observable in the center of
the Milky-Way Galaxy.

The long-term response of the star to tidal heating determines 
whether the star survives the encounter.  To be able to
survive the tidal circularisation near $r_t$, a star on an initially
wide orbit has to dissipate
\begin{equation}\label{eq:E}
   E_t \sim \left({\Mbh\over\Ms}
            \right)^{2/3}
            {G\Ms^2\over \Rs},
\end{equation}
which generally is much larger than the binding energy of the star.
As a consequence, a star can only survive a strong tidal interaction
if it cools efficiently. The most obvious way this can be achieved
is by radiation, i.e: a ``hot squeezar''.

The structure of a ``hot squeezar'' is unaffected by the encounter,
and as a result the cooling of the star is limited by the Eddington
luminosity.  The limiting luminosity can now be used to compute a
lower limit on the pericenter for which the star is able to survive an
encounter. The further orbital evolution can be described analytically
and the time-scale for complete circularisation of the orbit is then
(Alexander \& Morris \cite{AM03}):
\begin{equation}\label{eq:t0}
   t_0(r_p, a) = \frac{2\pi \Ms \sqrt{G M_\bullet a}}{\Delta E_t(r_p) }.
\end{equation}

If this time-scale is sufficiently small, the orbit of the star is not
changed significantly by scattering with field stars during inspiral
(section \ref{ssec:capture}). In this case the star circularises near
the tidal radius of the IMBH. The subsequent evolution of the IMBH
with main-sequence star binary is discussed in section
(\ref{ssec:bin}).

\section{Analytical estimates of tidal inspiral in presence of scattering}\label{sec:analytical}

If a star has a orbit in the tidal capture regime ($r_p\gtrsim r_t$)
after a first encounter with an IMBH, it experiences repeated tidal
interactions through which the orbit shrinks and and becomes
circularised.  The time-scale for circularisation is much larger than
the orbital period $P$ of the star ($t_0 \gg P$).
If the orbit is perturbed by another star during the inspiral process,
the inspiraling star may either be scattered to a wider orbit or 
into a tighter orbit.  In the first case tidal heating will become
less effective and the inspiral process slows down or stops
completely, making the star more vulnerable for subsequent dynamical
interactions. In the latter case the inspiral process is either accelerated
or, if the pericenter distance becomes smaller than the tidal radius of the star, 
the star is tidally disrupted by the IMBH and the inspiral process ends.

\subsection{Assumptions}\label{ssec:assump}

We present an approach which is quite similar for tidal disruptions and
tidal inspirals. To obtain analytical estimates for the two processes,
we make several assumptions that are
generally made for the treatment of tidal disruptions
(e.g. Lightman and Shapiro \cite{LS77}; Frank \& Rees \cite{FR76};
Cohn \& Kulsrud \cite{CK78}; Syer \& Ulmer \cite{SU99}; Magorrian \&
Tremaine \cite{MT99}; Miralda-Escud\'{e} \& Gould \cite{MG00}; Freitag
\cite{FR01a}; Alexander \& Hopman \cite{AH03}; Wang \& Merritt
\cite{WM04}) and inspiral processes (e.g. Hils \& Bender \cite{HB95};
Sigurdsson \& Rees \cite{SR97}; Ivanov \cite{IV02}; Freitag
\cite{FR01b}, \cite{FR03}; Alexander \& Hopman \cite{AH03}; Hopman,
Portegies Zwart \& Alexander \cite{HPZA04}; Hopman \& Alexander
\cite{HA05}).

Our simple analytical model captures some of the physical mechanisms
of the cluster and gives approximately the correct estimates for the
probability that a star is captured by the IMBH. Since many of the
assumptions listed here have been made by numerous previous papers, it
is of importance to trace to what extend the analytical model agrees
with the simulations, which has less number of simplifying assumptions.

Here we list our assumptions and discuss them.

\begin{enumerate}

\item The stellar distribution function is well approximated by a
 stellar cusp, i.e.

\begin{equation}\label{eq:BW}
  n_\star(r) = \frac{(3-\alpha)N_{a}}
                    {4\pi r_{a}^{3}}
               \left(\frac{r}{r_{a}}
	       \right)^{-\alpha}\,,
\end{equation}
where $\alpha$ is approximately constant within the radius of
influence $r_a$ of the IMBH (Bahcall \& Wolf \cite{BW76},
\cite{BW77}), and $N_a$ is the number of stars within $r_a$.

This assumption is shown to be satisfied by earlier N-body simulations
(Baumgardt et al. \cite{Baum04a}; \cite{Baum04b}; Preto, Merritt \&
Spurzem \cite{P04}), and also in our current simulations. The slope of
the cusp in our simulations is $\alpha\approx1.5$. A stellar cusp with
$\alpha\approx1.5$ has also been observed by star counts near the MBH
in our Galactic Centre (Alexander \cite{A99}). We will assume
$\alpha=3/2$ in the following analysis. This simplifies the
expressions somewhat, because in that case the relaxation time is
constant (see eq. [\ref{eq:tr}]).

Baumgardt, Makino \& Ebisuzaki (\cite{Baum04a}) have shown that for
small-mass black holes, containing less than a few percent of the
total cluster mass, the radius of influence of the black hole is given
by
\begin{equation}
r_a={G\Mbh\over\sigma_c^2},
\end{equation}
where $\sigma$ is the velocity dispersion in the cluster core. 

\item Within $r_a$ the orbits of the stars are Keplerian to good
approximation. This is a good assumption for our purposes, and it makes
many of the expressions more transparent.

\item The relaxation time within the cusp is much larger than the
    orbital time. Stars exchange energy and angular momentum only
    through small angle two-body interactions.

\item Within the radius of influence the stellar population can be
approximated by a single mass population. 

This assumption is made for simplicity, and much of the discrepancy
between the analytical model and the simulations stems from this
assumption. In a young stellar cluster there is a wide range of
masses, with a few stars which have masses much larger than the mean
stellar mass $\langle\Ms\rangle$. As a result the most massive stars
sink to the IMBH, and the masses within the radius of influence vary
strongly. This has considerable consequences for the dynamical
behaviour close to the IMBH which the simple analytical model fails to
describe. We will comment on this when we discuss the results of our
simulations.

\item Stars reach the IMBH by diffusion of angular momentum rather than
energy.

This assumption is accurate for a single mass distribution (Bahcall \&
Wolf \cite{BW76}). When a few very massive stars are present in the
cluster, however, dynamical friction operates on a time-scale $t_{\rm
d.f.}\sim (\langle\Ms\rangle/\Ms)t_r$, much shorter than the
relaxation time $t_r$. The dynamical mechanism which drives stars to
inspiral orbits in a young stellar cluster therefore typically
operates in two phases. First there is an ``energy phase'', during
which a massive star loses energy to field stars and sinks to the
centre; this is a purely elastic phase during which there is no tidal
heating. Then follows an ``angular momentum phase'', during which more
interactions with cluster stars change the angular momentum of the
massive star, until it finally reaches an orbit with pericenter close
enough to the tidal radius, and tidal heating becomes efficient.

\end{enumerate}

As a result of our simplifying assumptions, the following analysis
does not describe the radial distribution function (DF) of the stars
as a function of their mass. But once this DF is given, it does
provide a treatment of the interplay between scattering and
dissipation which is in good agreement with the simulations.

\subsection{Tidal disruption}\label{ssec:disr}

Tidal disruption of stars by an IMBH occurs when a star on an
eccentric orbit has angular momentum smaller than the
angular momentum at the ``loss cone'' ($J_{lc}$). For a main-sequence
star orbiting an IMBH, the loss-cone is defined as:
\begin{equation}\label{eq:Jlc}
   J_{lc} = \sqrt{2G\Mbh r_t}.
\end{equation}

The disruption rate and distribution of stars in angular momentum were
studied by Lightman \& Shapiro (\cite{LS77}). Here 
the average changes in momentum are much smaller
than $J_{lc}$ (diffusive regime).
For this case the time-scale for changes of the pericenter by
(small angle) scattering is (Alexander \& Hopman \cite{AH03})
\begin{equation}
\label{eq:tp}
   t_p(r_p, a)= \frac{r_p}{a} t_r.
\end{equation}
Here $t_r$ is the relaxation time, which we can estimate with
\begin{equation}\label{eq:tr}
   t_r=A_\Lambda \left({\Mbh\over\Ms}
                 \right)^2 
		 {P \over N(<r)}
\end{equation}
Here $A_\Lambda$ is a constant which includes the Coulomb logarithm,
and $N(<r)\propto r^{3-\alpha}$ is the number of stars enclosed within
$r$. For $\alpha=3/2$ the relaxation time is independent of distance
from the IMBH.

The diffusive regime is then defined to be the region where $a<r_{\rm
crit}$; $r_{\rm crit}$ is the semi-major axis for which $P(r_{\rm
crit})=t_p(r_t, r_{\rm crit})$,
\begin{equation}\label{eq:rcrit}
   r_{\rm crit} = \left( {1\over2\pi} \sqrt{G\Mbh} r_t t_r
                  \right)^{2/5}.
\end{equation}
The angular momentum of stars with $a<r_{\rm crit}$ changes by an
amount smaller than $J_{lc}$ per orbit, while the change in angular
momentum per orbit is larger than $J_{lc}$ if $a>r_{\rm crit}$. In the
latter case the velocity distribution can be isotropic, since it is
not affected by the presence of the loss-cone. This is not true for
the diffusive regime.

The distribution function and merger rate in the diffusive regime
are computed by solving the appropriate Fokker-Planck equation.  The distribution
function should become isotropic for $J\gg J_{lc}$ and vanish at the
loss-cone.  The steady state solution gives rise to a flow of stars
through the loss-cone, which is a consequence of the presence of a
mass sink at $J_{lc}$. This is reflected in the boundary condition
that the distribution function vanishes at the loss-cone.

The disruption rate in the diffusive regime is given by
\begin{equation}
  \Gamma_{d}= \int_{0}^{r_{\rm crit}}  
                \frac{ \mathrm{d}aN_{\mathrm{iso}}(a)}
		     {t_r\ln(J_{m}/J_{lc})}\,
            \approx \frac{(r_{\rm crit}/r_a)^{3/2}N_a}
                         {t_r\ln(J_{m}/J_{lc})}.
\label{eq:prompt}
\end{equation}
(e.g. Lightman \& Shapiro \cite{LS77}; Hopman \& Alexander
\cite{HA05}). Here $N(a)da\propto a^{2-\alpha}da$ is the number of stars
with semi-major axes in the range $(a, a+da)$. In eq. (\ref{eq:prompt})
$J_m$ is to be evaluated at $r_{\rm crit}$.

Equation (\ref{eq:prompt}) is interpreted as a slow diffusion of the
angular-momentum of the stars. The time-scale for the angular momentum
to change by a factor $\sim J_m=\sqrt{G\Mbh a}$ is of the order of the
relaxation time $t_r$. In a time $t_r$ a significant fraction of the
stars within $r_{\rm crit}$ are therefore disrupted by the IMBH. The
disruption rate is suppressed by a logarithmic factor which reflects
the presence of the loss-cone. This leads to a dilution of stars with
small angular momenta in the diffusive regime, as can be seen from the
solution of the Fokker-Planck equation (Hopman \& Alexander
\cite{HA05}).

For the tidal disruption rate, there is also a contribution from stars
for which the change in angular momentum per orbit is larger than
$J_{lc}$ (i.e., the regime $a>r_{\rm crit}$). In that case the
disruption rate $\Gamma(<a)$ for stars with semi-major axis smaller
than $a$ is estimated to be given by the fraction of stars in the
loss-cone, divided by the period,

\begin{equation}\label{eq:kick}
  \Gamma_d^{\rm kick}(<a)\sim{N(<a)r_t\over a P(a)}.
\end{equation}

For tidal capture (see below) there is no such contribution, and we do
not further consider stars in this ``kick'' regime here; for a
discussion see, e.g., Lightman \& Shapiro (\cite{LS77}); Cohn \& Kulsrud
(\cite{CK78}); Syer \& Ulmer \mbox{(\cite{SU99})}; Magorrian \& Tremaine
(\cite{MT99}).

\subsection{Tidal capture}\label{ssec:capture}

Due to relaxation in angular momentum, any star will eventually be accreted 
by the IMBH, either because of direct tidal disruption, or due to
tidal heating and inspiral.  If the number of stars in $(a, a+da)$ which
spiral in is given by $N_i(a) da$, and the number of stars which are
directly disrupted by $N_d(a) da$, then the probability that the star
reaches the IMBH by inspiral is given by

\begin{equation}\label{eq:Sa}
S(a) = \frac{N_i(a)}{N_i(a)+N_d(a)}
\end{equation}

Hopman \& Alexander (\cite{HA05}) performed Monte Carlo simulations to
show that stars with high energies (small semi-major axes) are
unlikely to enter the IMBH without first spiralling in, while
stars with low energies (large semi-major axes) are 
tidally disrupted without experiencing a tidal inspiral phase.

The rate of inspiral is then given by 
\begin{equation}
   \Gamma_{i} = \int_{0}^{\infty}\! \frac{\mathrm{d}aN(a)S(a)}
	                                 {t_r\ln(J_{m}/J_{lc})}.
\label{e:SU}
\end{equation}
Here the function $S(a)$ to accounts for the fact that
capture without premature disruption is improbable for $a>a_c$.

In order for the star to survive, tidal heating should happen at a
sub-Eddington rate (see previous section), yielding a minimal value
for the pericenter $r_{p, {\rm min}}\sim2r_t$. Inspiral without
premature scattering is only feasible provided that
$t_0(r_p,a)<t_p(r_p,a)$. As a consequence stars can only spiral in if
initially their semi-major axis is smaller than

\begin{equation}\label{eq:amax}
    a_{\mathrm{max}} = \left[\frac{3\Delta
        E_t(r_{p,\mathrm{min}})r_{p,\mathrm{min}}t_r}{2\pi \Ms
 \sqrt{G
        M_\bullet}}\right]^{2/3}.
\end{equation}
Note that $t_0>P$ and thus $a_{\rm max}<r_{\rm crit}$, which
implies that for tidal capture the relevant regime is the diffusive
regime, and that the tidal capture rate is smaller than the rate for
tidal disruption, in spite of the fact that tidal heating is efficient
for $r_p \lesssim 3r_t$, while tidal disruption happens only for
$r_p<r_t$.

The function $S(a)$ may be approximated by a Heaviside step-function
$\theta(a-a_c)$, where $a_c\gtrsim a_{\rm max}$ (Hopman \& Alexander
\cite{HA05}). This approximate behaviour is confirmed by our
simulations (see \S\ref{sec:results}). The rate at which stars
undergo inspiral due to tidal heating and survive is then given by
\begin{equation}
  \Gamma_{i}=\int_{0}^{a_{\rm
max}}\!\frac{\mathrm{d}aN(a)}{t_r\ln(J_{m}/J_{lc})}\,,\label{e:GULX}\end{equation}
or
\begin{equation}\label{eq:rate}
  \Gamma_i \approx \frac{(a_\mathrm{max}/r_a)^{3/2}N_a} {t_r\mathrm{ln}(J_{m}/J_{lc})}.
\end{equation}
Here the radial stellar distribution function is given by
Eq.\,\ref{eq:BW} with $\alpha = 3/2$.

Comparing Eqs.\,\ref{eq:rate} and \ref{eq:prompt} implies that the
rate at which stars in the diffusive regime are tidally disrupted is
larger than the tidal capture rate by 
\begin{equation}\label{eq:disr_cap}
   {\Gamma_d \over \Gamma_i} \approx \left({r_{\rm crit} \over a_{\rm max}}
                                     \right)^{3/2}.
\end{equation}

The actual tidal disruption rate is even higher, because stars with
semi-major axes $a>r_{\rm crit}$ can also be disrupted.

\section{Description of the $N$-body runs}\label{sec:Nbody}

We simulated the evolution of MGG-11 through $N$-body simulations of
star clusters containing $N = 131,072$ (128K) stars using Aarseth's
collisional $N$-body code NBODY4 (Aarseth \cite{AA99}) on the GRAPE6
computers of Tokyo University (Makino et al. \cite{MA03}). Following
Portegies Zwart et al.\ (\cite{PZea04}), our clusters are initially
King models with $W_0=9$ and half-mass radius $r_h=1.3$ pc. The
initial mass spectrum of the cluster stars was a Salpeter
mass-function between $1.0 M_\odot \le m \le 100 M_\odot$. These
cluster models have a projected half-mass radius, mass-to-light ratio
and total cluster mass after 12 Myrs that is consistent with the
properties of MGG-11 as observed by McCrady et al.\ (\cite{MC03}). The
clusters are also concentrated enough to form IMBHs via runaway
merging of stars. Stellar evolution was modelled according to Hurley
et al.\ (\cite{HU00}) and individual runs were calculated for $T=12$
Myrs. Table 1 gives an overview of the simulations performed. 

\input{table1}

Stars were merged if their separation became smaller than the sum of
their radii and the mass of the merger product was set equal to the
sum of the masses of the stars, i.e. we assumed no disruption of stars
and no mass loss during a collision. Since typical velocities of stars
in our clusters are much smaller than the escape velocities from
stellar surfaces, this should be a valid assumption.  The mass of
tidally disrupted stars was added to the mass of the central black
hole.

In total we performed 12 runs. In six of the runs we used a
relation for the radius of the runaway star derived from a fit to the
results of Ishii, Ueno \& Kato (\cite{IS99}) given by
\begin{equation}
 \log \frac{R_*}{R_\odot} = 1.5 \; \log \frac{M_*}{M_\odot} - 1.85.
\end{equation}
for stars with $120 M_\odot \le M_* \le 1000 M_\odot$. Radii of more
massive stars were set equal to the size of $10^3 M_\odot$
stars. These runs created fairly massive IMBHs with masses between
$3000 M_\odot < m < 4000 M_\odot$.  In another set of runs, we used
the mass-radius relation found by Bond, Arnett \& Carr (\cite{BO84}):
\begin{equation}
 \log {R_* \over R_\odot} =  0.47 \log \frac{M_*}{M_\odot} + 0.20. 
\end{equation}

While derived for stars with masses larger than $10^4 M_\odot$, this
relation gives radii in good agreement with the ZAMS radii from Hurley
et al.\ for stars with masses around $ 10^2 M_\odot$. These runs lead
to runaway stars with masses around $1000-2000 M_\odot$, roughly equal
to the estimated mass of M82 \mbox{ULX-1}. We did not apply stellar mass-loss
for the runaway stars and transformed them into IMBHs at $T=3$ Myrs.

\subsection{Implementation of tidal heating into the $N$-body simulations}

We have implemented orbital energy loss by tidal interactions between
a star and the IMBH in the $N$-body simulations using the prescription
of Portegies Zwart \& Meinen (\cite{PZ93}, henceforth PZM).  They
present fitting formula to the tidal energy loss as a function of the
masses, radii and polytropic indices of passing stars.  The
dimensionless functions $T_l(\eta)$ in eq.\ (\ref{eq:DEt}) for the
tidal energy were obtained from a polynomial fit to calculations by
Lee \& Ostriker (\cite{LE86}) and Ray et al.\ (\cite{RA87}) for stars
of polytropic indices $n=1.5, 2$ and 3.  PZM found that the $l=2$ and
$l=3$ terms give a 99\% contribution to the dissipated energy and
therefore did not take higher harmonics into account.

In our implementation, we assumed that main sequence stars with
masses $m<0.4 M_\odot$ and giants have polytropic index $n=1.5$,
while stars with mass $0.4 M_\odot < m < 1.25 M_\odot$ have polytropic
index $n=2$. More massive main-sequence stars were assumed to have
$n=3$. It was not necessary to apply tidal heating to white dwarfs or
neutron stars since these stars were not present in the runs because
the studied clusters are too young.

Eq.\ 3 is only valid if the orbital eccentricity is
close to 1. For orbits of smaller eccentricity the amount of tidal
heating is smaller since the star feels the influence of the IMBH also
at apocenter. In order to reflect this we multiply the heating energy
$\Delta E_t$ by the eccentricity $e$ of the orbit. In our current
implementation, the total angular momentum is conserved and an unperturbed
IMBH-star system will always circularise at twice the initial
pericenter distance $r_{pi}$.

Due to the structure of NBODY4, tidal heating had to be included in
three different places in the code. In the main integrator, we checked
the distance of each particle to its nearest neighbour each time it was
advanced in time. This is facilitated by the GRAPE hardware which
provides the index of the nearest neighbour for each particle at
almost no extra computational cost. The closest encounter distance was
calculated once the relative velocity $v_{rel} = \vec{r}_{ij}
\vec{v}_{ij}/|\vec{r}_{ij}|$ between two particles switched sign from
negative to positive and we applied tidal heating to all stars that passed
within $5 r_t$ of a massive black hole by decreasing the kinetic
energy of the combined star-black hole system by an amount $\Delta
E_t$. Energy was extracted by changing the velocity and position of the
black hole and the star when the system was at pericenter. The energy was changed 
according to eq.\ 3 such that the total linear and angular momentum remains
constant.

Since in NBODY4 the motion of close particle pairs is followed by KS
regularisation (Kustaanheimo \& Stiefel \cite{KU65}), we also had to
add tidal heating to the motion of KS binaries. For unperturbed pairs
this was done each time their motion was integrated by changing the
semi-major axis and eccentricity of the binary according to the
accumulated amount of tidal energy. For perturbed binaries, tidal
heating was applied after each pericenter passage since at this point
the perturbation of the motion of close neighbours is smallest and it
is easiest to correct the total energy of the cluster.

Energy dissipation by gravitational radiation was implemented
following Peters (\cite{PE64}) formulae in a similar fashion
as the implementation of tidal heating.
\begin{figure}
\epsfxsize=8.3cm
\begin{center}
\epsffile{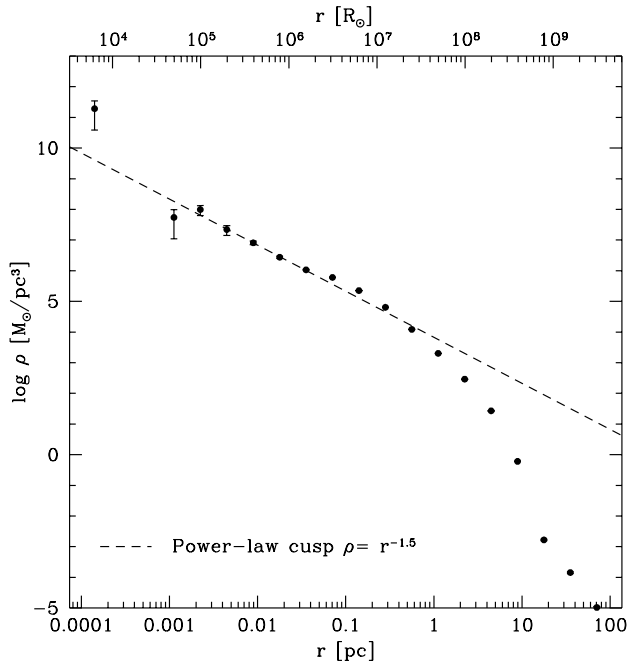}
\end{center}
\caption{Density profile of run 6 at $T=3.15$ Myrs, just before an
inspiral event happens.  Between $10^{-3} {\rm pc} < r < 0.05$ pc,
the density profile of the cluster follows a power-law cusp with
$\alpha=1.5$.  The innermost data point is due to one star which
experiences an inspiral at $T=3.15$ Myrs.}
\label{fig1}
\end{figure}

\section{Results}\label{sec:results}

\subsection{Central density and mass segregation}

We begin by discussing the density profiles of the star clusters after
an IMBH has formed in the centre.  Fig.\ 1 depicts the density profile
of run 6 from Table 1 at $T=3.15$ Myrs just before a tidal capture
event occurs. The cluster centre is assumed to coincide with the
position of the IMBH. It can be seen that the cluster follows a
power-law mass density profile $\rho \sim r^{-\alpha}$ with slope $\alpha =
1.50$ inside $r = 0.05$ pc.  This is close to the value of $\alpha =
1.55$ found by Baumgardt, Makino \& Ebisuzaki (\cite{Baum04b}) for
IMBHs in multi-mass clusters.  Inspection of the calculations shows
that a cusp is already present by the time the runaway star has
collapsed to a black hole (at about 3 Myrs), and as a result already
influences the late stages of runaway merging. One reason for the
quick formation of cusps is the large initial mass spectrum of our models which
reduces the relaxation time. Fig.\ 1 also justifies our assumptions
about the presence of a cusp around the IMBH in \S 3.
Due to the finite number of cluster stars, the central cusp runs out
of stars at $r \approx 10^{-3}$ pc, corresponding to $r = 10^5
R_\odot$. In the present example, one star was pushed to $r = 10^{-4}$ pc due
to dynamical interactions between the inner cusp stars. This star
experiences an inspiral event at $T=3.15$ Myrs.
\begin{figure}
\epsfxsize=8.3cm
\begin{center}
\epsffile{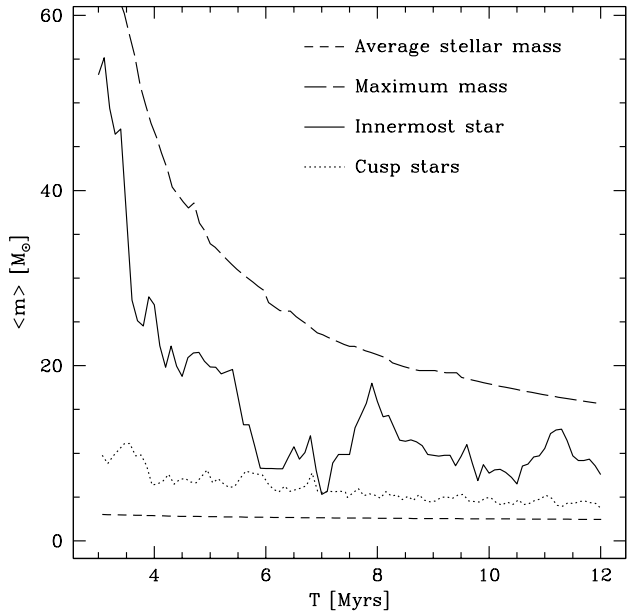}
\end{center}
\caption{Average and maximum masses of stars in the cluster as a
         function of time. The masses of all stars decrease as a
         result of stellar evolution. Due to mass segregation, stars
         in the cusp around the IMBH are more massive than the average
         cluster stars and the star closest to the IMBH is always one
         of the most massive stars in the cluster.}
\label{fig2}
\end{figure}

Fig.\ \ref{fig2} shows average and maximum masses of stars in the
cluster and near the IMBH over the course of a simulation.  Values
shown are averaged over several simulations. Masses of high-mass stars
decrease as a result of stellar evolution since stars
lose mass in winds and the most massive stars are constantly removed
and turned into compact remnants. As a result, the mass of the most
massive cluster star, which is near 80 $M_\odot$ at $T=3$ Myrs, drops
to 15 $M_\odot$ at the end of the simulation.  Due to the large number
of low-mass stars, the average mass of cluster stars stays nearly
constant within the first 10 Myrs at $\langle m \rangle \simeq 3
M_\odot$.  Due to mass segregation, the mean mass of stars in the cusp
is higher than the mean mass of all cluster stars.  Also within the
cusp, massive stars sink close to the IMBH so the stars near the IMBH
have masses significantly higher than the mean mass. This shows that
it is important to take mass segregation into account when determing
the inspiral rates.

\subsection{Tidal capture}

Several trends can be deduced from Table 1: 

\begin{itemize}

\item Successful inspiral events occur in all runs with on average $\sim 4$
inspiral events per run during the first 9 Myrs after IMBH formation. Hence there is a high chance that an
IMBH in a young star cluster like MGG-11 will undergo a RLOF phase at
least once during the lifetime of the cluster.

\item Since the masses of stars in the inner cusp are relatively large
(see Fig.\ \ref{fig2}), most inspiral events also involve massive
stars. The average mass of captured stars is $ \langle m \rangle
\approx 25 M_\odot$, which exceeds the average mass of stars in the
cusp. This is caused by the larger cross-section for a tidal
interaction with the IMBH of the massive stars.

\item In agreement with the theoretical estimates of \S 3, successful
tidal inspiral occurs only for stars in tightly bound orbits around
the IMBH with semi-major axis $a<10^5 R_\odot$, the exception being giant
stars with much larger stellar radii. This corresponds
roughly to the inner end of the cusp. Stars on orbits with
larger semi-major also experience tidal interactions with the IMBH
(see Fig.\ \ref{noinsp}), but are scattered away by cusp stars before their 
orbits can circularise.

\item Although the cross-section for tidal heating is larger than the
cross-section for tidal disruption, it is much more likely to tidally
disrupt stars, than to bring a star into a close orbit around the IMBH
via tidal heating. Averaged over all runs, we find that the ratio
of tidal disruptions to inspirals is $N_d/N_i \approx 7$. 

\item Successful inspiral events occur predominantly 
early on in the cluster evolution: Between $5< T < 12$ Myrs only 24 inspiral
events occur compared to 26 events between $3 < T < 5$ Myrs. 
The decay in the number of tidal inspiral events is mainly
caused by the expansion of the cluster, which is driven by mass
loss from stellar evolution.  In addition, the most massive cluster
stars leave the main sequence and increase their radii at $T \approx
3.3$ Myrs, which explains the large number of inspirals around this
time involving stars with masses $m > 25 M_\odot$.

\item The final pericenter distances after the orbit circularised,
are within a factor of three of the tidal radii 
since the timescale for tidal circularisation is small only for small
$r_p/r_t$. 

\end{itemize}

\begin{figure}
\epsfxsize=8.3cm
\begin{center}
\epsffile{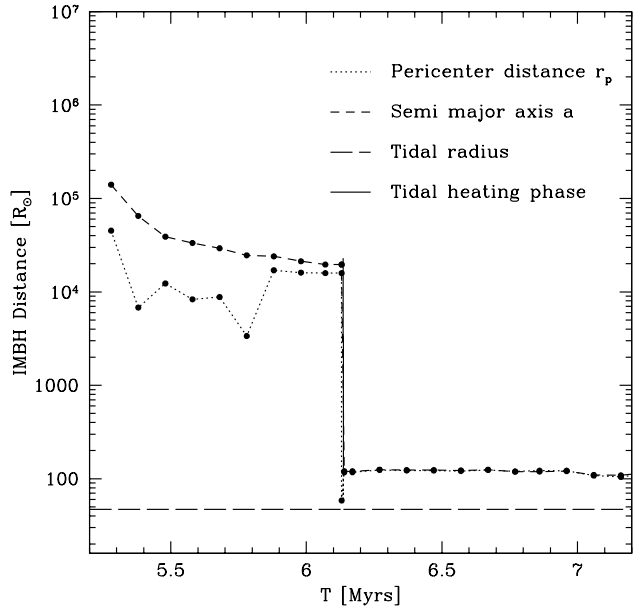}
\end{center}
\caption{Semi-major axis and pericenter distance of a tidally heated
star before and after inspiral due to tidal heating.  The
main tidal heating event happens at $T=6.13$ Myrs 
at which time the star
circularises. Later, the semi-major axis changes due
to external perturbations while the tidal radius increases as
a result of stellar evolution and IMBH growth.}
\label{insp}
\end{figure}

As an example for an inspiral event, Fig.\ \ref{insp} depicts the
evolution of semi-major axis $a$ and pericenter distance $r_p$ of a
captured $m=17.6$ $\mbox{M}_\odot$ star with time from one of
our runs (run 3). The time when tidal heating shrinks the orbit is
shown by a solid line. Between 5.3 and 6.1 Myrs the star segregates towards
the IMBH as a result of its higher than average mass and the semi-major axis of the orbit
shrinks to a few $10^4 R_\odot$.
Tidal heating occurs at $T=6.13$ Myrs, when the pericenter distance is scattered to a distance only
slightly larger than the tidal radius. 
During the inspiral the semi-major axis decreases from more than
$10^4$\,R$_\odot$ to about 120 $R_\odot$, and the orbit circularises
at $r_{pf}/R_t = 2.5$. After the tidal circularisation, the semi-major 
axis remains unchanged except for perturbations due to passing stars, which
increase the eccentricity. Upon each
time a small eccentricity is induced, tidal effects will directly start
to circularise the orbit again, and so result in a slight decay of the
semi-major axis.  A strong orbital perturbation causes the star to fill  
its Roche-lobe at $T \simeq 10.2$ Myr and it is destroyed by the IMBH.
Our present $N$-body code
does not allow to treat RLOF phases, a star whose stellar radius expands
or whose orbital radius shrinks to the point where the stellar radius
exceeds the tidal radius will simply be disrupted.  We will present a
discussion about the further
evolution of such systems based on simple binary-stellar evolution
estimates in chapter 5.5.

\subsection{Super-Eddington heating rates}\label{ssec:heat}

The inspiral process may lead to complete tidal circularisation, but
also to the destruction of the star. The latter happens if 1)
the star passes the black hole's event horizon or 2) if the rate of
energy dissipation in the tidal encounter exceeds the Eddington
luminosity.  We will now estimate the latter fraction.  For each
inspiral event we calculated the rate at which energy was pumped into the star 
due to tidal heating when the eccentricity of the orbit was $e=0.9$.
At this time the amount of
energy dissipation per time $\Delta E/P$ reaches a maximum since
$a$ and $P$ decrease with $e$ while for smaller eccentricities
the stars feel the influence of the black hole also
at apocenter and $\Delta E$ decreases.
\begin{figure}
\epsfxsize=8.3cm
\begin{center}
\epsffile{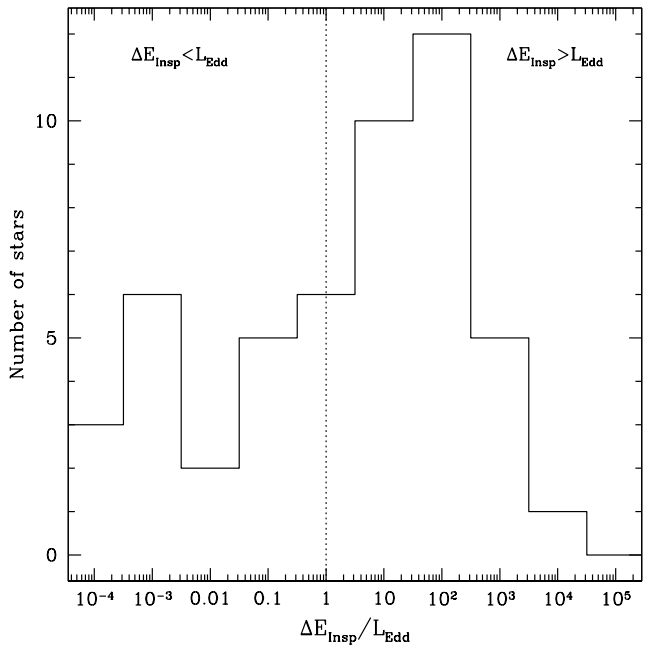}
\end{center}
\caption{Energy dissipation rates $\Delta E_{\rm Insp}$ when the
orbital eccentricities have reached $e=0.9$ compared with the
Eddington luminosities $L_{Edd}$ of the stars. Nearly half of
all stars experience sub-Eddington heating rates during the
inspiral process and survive the inspiral.  Moderately
super-Eddington stars might also survive the inspiral since
the super-Eddington heating rates occur for only short
intervals and the extra energy pumped into the star is not
sufficient to disperse it completely.  If the amount of tidal
heating is too large, the stars expand and are tidally
disrupted by the IMBH.}
\label{heat}
\end{figure}

Fig.\ \ref{heat} compares the energy dissipation rates at $e=0.9$ with
the Eddington luminosities of the stars. Super-Eddington heating rates
occur in roughly half the cases, but since Fig.\ \ref{heat} shows the maximum heating
rates during the inspiral which occur only for a few orbits, cases with
$\Delta E_{Insp}>L_{Edd}$
do not necessarily end up in the destruction of the star. Stars might for
example lose their outer envelopes and shrink, thereby
becoming less susceptible to tidal perturbations, so that the central parts
of the stars survive the inspiral. In addition, the extra energy pumped into the
star must be larger than the potential energy of the star in order to disrupt it,
so the super-Eddington heating rates have to be maintained over many orbits.
In order to answer the question which stars survive the inspiral, more detailed simulations
treating the response of the star to the tidal heating would be required.
These are beyond the scope of the present paper.

\subsection{Comparison with analytic estimates}

We compare the numerical results with the analytical estimates from \S
(\ref{sec:analytical}) for the following rather typical example:
$\Mbh=3\times10^3\Mo$, $\Ms=30\Mo$, $\Rs=10\Ro$, $t_{rc}=3\;{\rm
Myrs}$, $r_a=0.05\;{\rm pc}$, $N_a=300$, and $r_p/r_t=2.3$. With
eqs. (14), (19) and (21) we then find that over a time interval of
$T_{f}=9\;{\rm Myr}$ the number of tidal disruptions is $N_d=\Gamma_d
T_{f}=8.6$, while the number of tidal captures is smaller by a factor
$(r_{\rm crit}/a_{\rm max})^{3/2} \simeq 0.08$, $N_i=\Gamma_i T_{f}
\simeq 0.69$, i.e: $N_i/N_d \simeq 0.08$.  The maximum distance from
which stars can originate is $a_{\rm max} \simeq 3.4\times10^4\Ro$. By
using eq. (\ref{eq:rcrit}), we also find that the critical radius
inside which the loss cone should become empty is $r_{\rm crit} = 2
\cdot 10^5 R_\odot$, slightly larger than the inner end of the
cusp. 

From our simulations we find that $N_d = 27$, $N_i = 4.1$ and
$N_i/N_d = 0.15$.  Using only the inspirals which have sub-Eddington
heating rates gives $N_i = 1.58$ and $N_i/N_d = 0.06$
Hence the results for the analytical inspiral and disruption rates
$N_i$ and $N_d$ are within a factor of 3 to those found in the
simulations while the ratio $N_i/N_d$ is correctly predicted.
\begin{figure}
\epsfxsize=8.3cm
\begin{center}
\epsffile{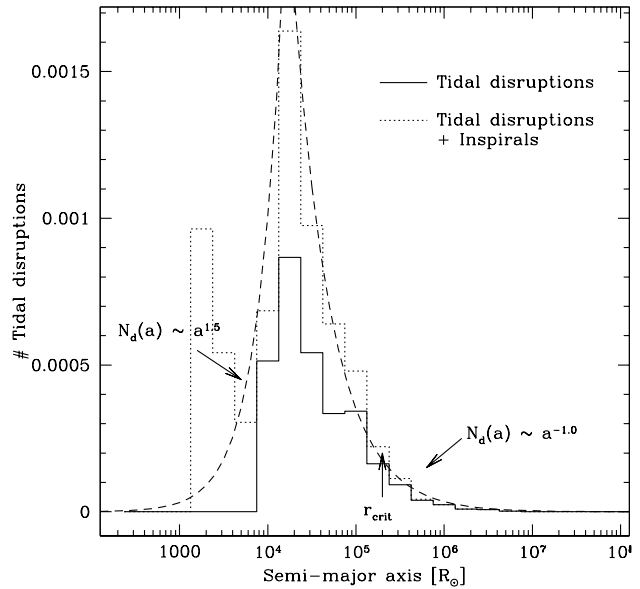}
\end{center}
\caption{Initial semi-major axis distribution of tidally disrupted
stars (solid lines) and combined distribution of tidally
disrupted and captured stars (dotted). Shown are differential
rates $N(a)=(N(>a)-N(>a+\Delta a))/\Delta a$.  The combined
distribution has a maximum near the critical radius $r_{\rm
crit}$ where the loss-cone becomes empty, in agreement with
theoretical predictions (Lightman \& Shapiro 1977).  The
combined distribution can be fitted by two power-laws.}
\label{disr1}
\end{figure}

Although tidal disruptions are not the main focus of the present
paper, it is interesting to compare them with analytic estimates since
the physical processes responsible for inspirals and disruptions are
the same. The distribution of semi-major axis of tidally disrupted
stars has been calculated by Lightman \& Shapiro (\cite{LS77}).  They
found that the peak in semi-major axis distribution should occur near
$r_{\rm crit}$ (see eq. [\ref{eq:rcrit}]). Inside $r<r_{\rm crit}$,
the loss cone is empty, so stars have to be scattered into the loss
cone by two-body relaxation. In this case the number of disrupted
stars should be proportional to $N_d(<r) \sim N(<r)/t_r$ which leads
to a scaling $N_d(<a) \sim a^{3/2}$ for the number of encounters with
semi-major axis smaller than $a$ (see eq. [\ref{eq:prompt}]). For the
corresponding differential function we find therefore $N_d(a) \sim
a^{1/2}$.  At larger distance $r>r_{\rm crit}$ the loss cone is full
(``kick'' regime); the number
of disrupted stars is then given by eq. (\ref{eq:kick}). This leads to
a scaling $N_d(>a) \sim a^{-1}$, or, for the differential function
$N_d(a) \sim a^{-2}$.
\begin{figure}
\epsfxsize=8.3cm
\begin{center}
\epsffile{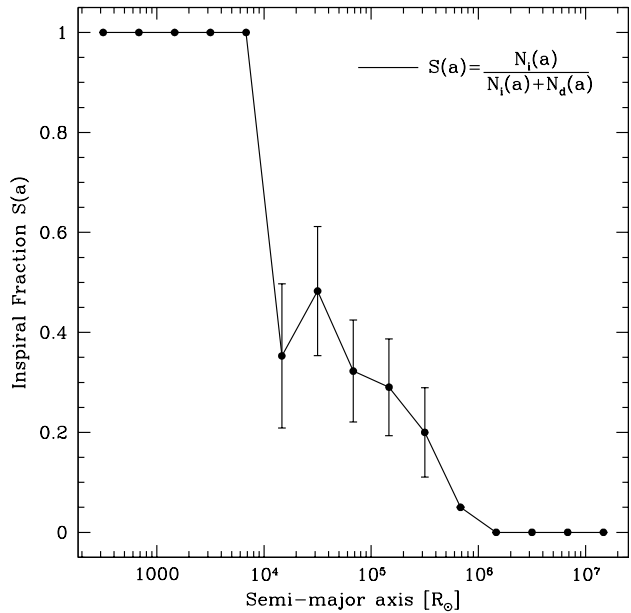}
\end{center}
\caption{Fraction of tidally captured stars compared to all stars
         ending in or near the IMBH as a function of the initial
         semi-major axis $a$ of the stars before tidal
         capture/disruption (see $S(a)$ in eq. [\ref{eq:Sa}]).  For
         small semi-major axis tidal capture dominates while for
         larger radii tidal disruption becomes the dominant process
         for feeding the central black hole since stars cannot spiral
         in any more before being scattered away by other stars.}
\label{disr2}
\end{figure}

Fig.\ \ref{disr1} shows the number of circularised and tidally
disrupted stars as a function of the initial semi-major axis.  The
maximum occurs at $a_{\rm peak} \simeq 2\cdot10^4 R_\odot$, somewhat 
smaller than our analytical estimate for the critical radius $r_{\rm crit}$.

For radii smaller than $a_{\rm peak}$, the combined distribution of
tidally heated and disrupted stars can be fitted by a relation $N(a) \sim
a^{1.5}$. This is somewhat steeper than predicted by Lightman \& Shapiro
(\cite{LS77}), however the uncertainties in the observed slope are quite
large due to the small number of events at small radii. If real, the reason for the
discrepancy with Lightman \& Shapiro could
be the capture of stars by the IMBH followed by three body encounters with
other stars in the cusp. An inspection of the output data shows that
these processes contribute considerably to the tidal destruction
rate but are not taken into account by our analytic estimates.
For radii larger than $a_{\rm peak}$, we obtain a slower decrease than
predicted by Lightman \& Shapiro (\cite{LS77}). The reason could
be that the stars in the cusp drive the Brownian
motion of the IMBH. This influences
the determination of the semi-major axis of in-falling stars in case
of large semi-major axis.

Hopman \& Alexander (\cite{HA05}) found that if stars move slowly
through phase space,
the number of tidally captured stars should be much larger
than the number of tidally disrupted stars for small radii, $N_i(a)\gg N_d(a)$; while
for large radii two-body relaxation becomes important and prevents
tidal inspiral so $N_d(a)\gg N_i(a)$. In Fig.\
\ref{disr2} we compare the number of stars that are tidally heated
with all stars that strongly interact with the IMBH.  It can be seen
that we indeed find the predicted behaviour for S(a). 

\begin{figure}
\epsfxsize=8.3cm
\begin{center}
\epsffile{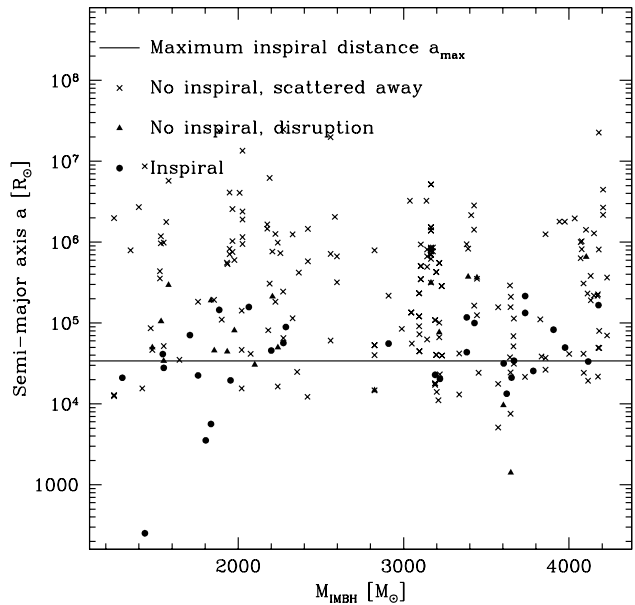}
\end{center}
\caption{Semi-major axis of tidally circularised stars (dots) compared
         to stars for which tidal heating does not lead to
         circularisation (crosses). Successful tidal inspiral is only
         possible for stars on orbits with small semi-major axis. The
         threshold between the two cases is in agreement with the
         prediction of this paper and of Hopman et al. (2004) for the
         maximum inspiral distance $a_{\rm max}$.} 
\label{noinsp}
\end{figure}

In \S 3.3  we demonstrated that the semi-major axis of a stellar orbit
has to be smaller than $a_{\rm max}$ in order for the orbit of the star 
to be successfully circularised. Otherwise the star is scattered away from
its orbit due to encounters with other stars.
Fig.\ \ref{noinsp} shows the initial semi-major axis of stars
which are tidally heated by the IMBH. Stars for which the inspiral
continues all the way until the orbit is circularised are shown by
filled circles while stars for which the heating process is terminated
since they are either scattered away or experience super-Eddington
heating rates during the inspiral process are shown with crosses and
triangles respectively. 

Tidal capture is only successful if the initial semi-major axis of
the star $a \aplt 2 \cdot 10^5 R_\odot$. The average inspiral distance
agrees quite well
with the predicted value of $a_{\rm max}$. Stars circularising
due to tidal
heating fall into two categories: most of them initially moved on
orbits with $e \approx 0.99$ and come from the inner cusp, while few
came from relatively low-eccentricity orbits with semi-major axis of a
few $1000 R_\odot$.  Most of the stars originating from low-$a$
orbits experienced dynamical
interactions with previously inspiraled stars. 

\subsection{Further Evolution of captured Stars}\label{ssec:bin}

In order to study the mass accretion rate and X-ray luminosity of the
IMBHs, we assume that only the innermost stars are close enough
to transfer a significant amount of mass to the IMBHs and use the model 
discussed by K\"ording, Falcke \& Markoff
(\cite{KFM02}). They assume that X-rays are created in accretion discs
and jets and argue that a black hole with a nearby companion is in the
hard state if $\dot{M} > \dot{M}_{\rm crit}$ in which case the disc
luminosity is given by $L_X=\epsilon \dot{M} c^2$. At lower accretion
rates $L_X=\epsilon \dot{M}^2 c^2/\dot{M}_{\rm crit}$. Following their
paper, we adopt $\epsilon=0.1$ and $\dot{M}_{\rm crit}=0.1
\mbox{M}_\odot/$Myr. The mass-loss rates of the donor stars are
directly taken from the stellar-evolution routines of Hurley et al.\
(\cite{HU00}). 
Stellar winds have typical velocities of about 200 km/sec for
LBVs and 1000 km/sec for Wolf-Rayet stars (Paumard et al. 2001), so
depending on the orbital separation of the innermost star and the IMBH, gas lost
by the star can also escape from the system instead of ending up on the
IMBH. A detailed study of gas accretion would require SPH calculations
like the ones presented by Cuadra et al. (\cite{Cea05}), which are
beyond the scope of this paper. In order to account for the loss of
gas, we have plotted times when the escape velocity of the star is
smaller than 200 km/sec with crosses. In these cases the values for
$L_X$ should be regarded as upper limits.
\begin{figure}
\epsfxsize=8.3cm
\begin{center}
\epsffile{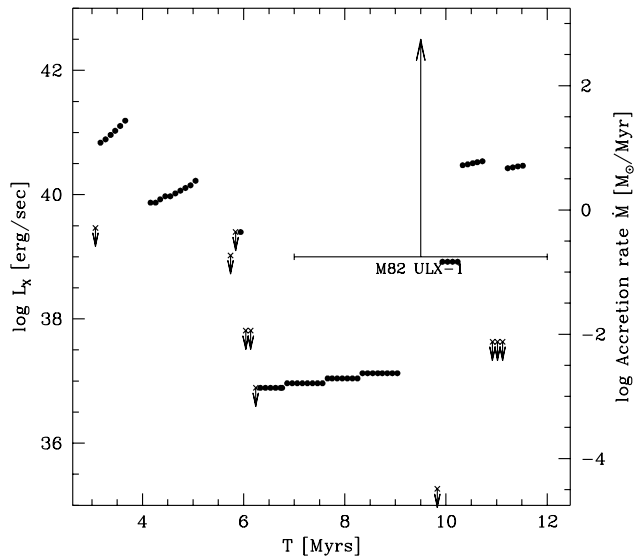}
\end{center}
\caption{Mass accretion rate and X-ray luminosity of the IMBH of run~10.
Crosses denote times when the distance of the star closest to
the IMBH is so large that most of the mass lost from the star
escapes from the star-IMBH system. Between 3 and 12 Myrs, several
close companions to the IMBH formed through tidal heating which
transfer mass to the IMBH.  For two events, the X-ray luminosity of
the IMBH reaches $L_X=10^{40}$ erg/sec while the stars are on the
main-sequence, enough to explain the X-ray luminosity of M82 ULX-1
and also within the right age range of MGG-11.}
\label{xray}
\end{figure}

\begin{figure}
\epsfxsize=8.3cm
\begin{center}
\epsffile{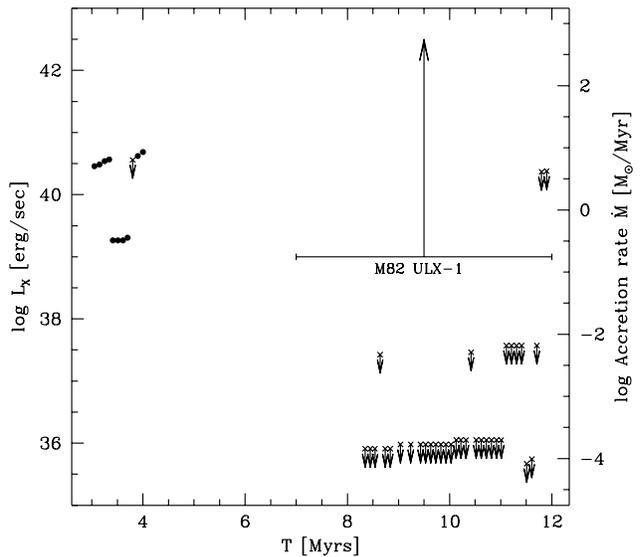}
\end{center}
\caption{Same as Fig. \ref{xray} but now for run 7. This cluster also experiences several inspiral events between 3 and
12 Myrs, but none of them produces an X-ray source which is in the right age range and bright enough
to explain M82 ULX-1.}
\label{xray2}
\end{figure}

Figs.\ \ref{xray} and \ref{xray2} depict
the X-ray luminosities of the IMBHs as a function of time for runs 7
and 10.
For both clusters, the distance and nature of the innermost star changes rapidly during the simulation as
stars near the IMBH are turned into compact remnants by stellar evolution and then scattered 
out of the cusp by other stars
or are tidally disrupted by the IMBH. Run 10 experiences 4 prolonged ULX phases between 3 and 12 
Myrs, each time due 
to massive main-sequence stars brought close to the IMBH as a result of tidal heating. The last 
two phases are within
the right age limits for MGG-11 as determined by McCrady et al. and produce X-ray sources with 
a luminosity exceeding
$10^{40}$ ergs/sec. Run 10 would therefore be a good candidate to explain the nature of the 
ULX in M82.

In run 7, three stars experience tidal inspiral between $3 < T < 4$
Myrs and produce ULX sources before the stars are tidally disrupted by
the IMBH or turned into compact remnants. At later times, the
innermost stars are always too far away from the IMBH for significant mass
transfer, hence run 7 is not able
to explain the ULX in M82. 
In total we find that in 10 of the 12 performed runs a ULX source is produced at 
least once between
3 and 12 Myrs and that 4 out of the 12 runs create an X-ray source
brighter than $2\cdot10^{39}$ erg/sec within the age range of
MGG-11. The average time an X-ray source with a luminosity brighter
than $10^{39}$ erg/sec is present in our runs is 0.96 Myrs, while
a source brighter than $10^{40}$ erg/sec is present for 0.62 Myrs. 
These values would be lower if tidal destruction of stars with
super-Eddington heating rates would be taken into account (see section \ref{ssec:heat}), 
however, at the moment our code does not follow the RLOF phases which would
significantly prolong the lifetime of the ULX sources and increase
their luminosities. We therefore conclude that an IMBH in a dense 
star cluster like MGG-11 has a high chance of creating an ULX and
that our model can explain the ULX source in M82. 

\section{Conclusions}\label{sec:concl}

Observations of X-ray sources with luminosities higher than the
Eddington luminosity of stellar mass objects provide strong support 
for the existence of IMBHs. However, the mere existence of
an IMBH does not guarantee the existence of ULXs; there has to be a
mechanism which accounts for the accretion of gas by the IMBH. 

In this
paper we confirm the hypothesis that massive ($\Ms>1000\,\Mo$) stars,
which may be the progenitors of IMBHs, form naturally in young dense
clusters as a result of many collisions between young stars. We assume
that the massive star indeed forms an IMBH, and we perform N-body
simulations of the interaction between the IMBH and its host cluster,
while accounting for tidal encounters between stars and the IMBH. 
Madhusudhan et al.\ (\cite{M05}) 
found through stellar evolution studies that massive main-sequence stars
with masses $\Ms > 8 M_\odot$ in orbits of 6 to 30 times the radius of
the donor star around an IMBH are necessary to explain ULX sources.
Our simulations show that tidal heating of stars in young star clusters will
create exactly such systems. We
find that, as a result of tidal heating, stars are likely to
circularise at distances of a few tidal radii from the IMBH. Once the
star has circularised, accretion of stellar gas by the IMBH is
sufficient to account for luminosities as high as $L_x \apgt
10^{39}$\,erg/s. On average, the X-ray luminosities of the circularised
stars exceed $10^{39}$ ergs/sec for almost $10^6$ yrs within the first
12 Myrs in our runs, adding further credibility to the scenario that
ULX are IMBHs accreting gas lost from nearby companion stars.

Our conclusions differ from those reached by Blecha et al. (\cite{B05}). The main reason could
be that the IMBHs in our runs are more massive than the
ones they considered and they found that higher-mass IMBHs have a higher chance
of capturing companions. In addition, they also neglected the effect of
mass segregation, whereas in our runs we find that mainly massive stars circularise
around the IMBHs. 

As we noted in \S (\ref{sec:results}), stars which are tidally captured
by an IMBH are typically among the most massive stars in the
cluster. 
The massive star loses much of its energy to other stars by dynamical
friction. In a single mass distribution (as is often assumed in an
analytical treatment of the problem), diffusion in energy space is
very slow (Bahcall \& Wolf \cite{BW76}), but when a spectrum of masses
is present, energy loss due to dynamical (non-elastic) interactions
with field stars (as opposed to tidal interactions with the IMBH)
plays an important role. This further emphasises the
importance of mass segregation.

In spite of the fact that some highly simplifying assumptions were
made, the final estimate for the rate resulting from our analytical
treatment is of the same order of magnitude as what we find from our
simulations.

After the donor star leaves the main sequence, it forms a
compact remnant. For high-mass donor stars the remnant will be a
stellar mass BH and the subsequent evolution of the IMBH-BH binary is
driven by the emission of gravitational waves until eventually the
binary members merge. Hopman \& Portegies Zwart (\cite{HPZ05})
have shown that the event rate for this is likely to be high enough to be
detectable by LISA, in particular if the IMBH mass is larger than
$\sim3\times10^3\;\Mo$. Observations of gravitational waves from such a
binary will give further support for the scenario discussed in this
paper. In addition,  if IMBHs are formed in star clusters near the centres
of galaxies and spiral in due to dynamical friction on the field stars,
mergers of the IMBHs with the MBHs at galactic centres would also be important  
{\it LISA} sources (Portegies Zwart et al.\ \cite{PZea05}).

The mechanism of stellar capture by gravitational wave radiation is
similar to that of tidal capture. Mass segregation probably plays a
less significant role in the case of the inspiral due to gravitational
wave emission by compact remnants when spiralling in to a MBH of
$\Mbh>10^6\Mo$, because in that case the mass ratio of the field stars
is less extreme than is the case for a young dense stellar
cluster. The number of compact remnants in galactic centres is not
known. Observationally the mass of these objects can be constrained in
our Galactic Centre by finding deviations from pure Keplerian orbits,
in particular pericenter-shift (Mouawad et al. \cite{MO04}). Current
observational constraints are not conclusive, but this situation will
improve with more accurate measurements of the stellar orbits in our
Galactic Centre. 

In addition to direct disruption and tidal inspiral, we also find a large number of
stars that experience one or several strong tidal encounters, but are
scattered to wider orbits before they can circularise near the IMBH. As a
result of the strong tidal encounters, such tidally heated stars or
``tidally scattered'' stars are expected to show signs of mixing,
large spin and stripping; this may be directly observable in the Galactic
centre (Alexander \& Livio \cite{AL01}). We found that on average
there are 45 stars
which have one or more interactions of less than $5r_t$ with the MBH and are
then scattered away. We expect that this number scales linearly with
pericenter distance.

We have assumed that the IMBHs in the clusters formed in a runaway
merger of young stars, during a very early stage of cluster evolution.
However, other scenarios of IMBH formation have been discussed in the
literature. IMBHs may form as remnants of population III stars
(e.g. Madau \& Rees \cite{MR01}), or as the merger of stellar mass
black holes due to GW emission (Miller \& Hamilton \cite{MH01}). If
IMBHs indeed form in these scenarios, and they still reside in stellar
clusters, the stars in these clusters will have a much lower maximal
mass than in the young clusters we discussed here. If a star is
tidally captured, accretion will most likely not lead to high
luminosity to account for the most luminous ULXs, although
accretion via the subsequent RLOF will lead to a low-luminosity
X-ray source.

\section*{Acknowledgements}
We are gratefull to Sverre Aarseth for helping us with
NBODY4, and to Tal Alexander for interesting discussions.  This work
is supported by Minerva grant 8484, the Dutch Organization for
Scientific Research (NWO under grant \# 635.000.001), the Dutch
Advanced School for Astronomy (NOVA), and the Royal Netherlands
Academy of Arts and Sciences (KNAW).  We thank the Sternwarte Bonn
and the University of Amsterdam for their hospitality.

\bsp

\label{lastpage}

\end{document}

%% file: table1.tex
\begin{table*}
\caption[]{Details of the performed $N$-body runs. Shown
are the mass-radius relation assumed for the runaway stars (Ishii et
al. or Bond et al.), the mass $M_{\bullet\;i}$ of the IMBH at time of
its formation (T=3 Myrs), the mass $M_{\bullet\;f}$ of the IMBH at the
end of the run (T=12 Myrs) and the number $N_{Dis}$ of stars tidally
disrupted by the IMBH. Final columns give orbital parameters for 
tidal heating events: Starting time $T_{\rm Insp}$ of the 
inspiral, duration of inspiral $\Delta T_{\rm Insp}$ until the orbit
is circularised, mass $\Ms$ and radius $\Rs$ of the star at time
of inspiral, tidal radius $R_t$ of the star, initial
pericentre distance of the orbit $r_{\rm p, i}$, ratio of
pericentre distance to the tidal radius of the star $r_{\rm p,
i}/r_t$, semi-major axis $a$ and eccentricity $e$ of the initial
orbit and semi-major axis $r_{\rm p, f}$ after the orbit
circularised.
}
\begin{tabular}{c@{}c@{}c@{}c@{}c@{}rcr@{$\;\;$}r@{$\;\;$}r@{$\;\;$}rcrrr}
\noalign{\smallskip}
 Run & M-R & \multicolumn{1}{c}{$M_{BH\;i}$}& \multicolumn{1}{c}{$M_{BH\;f}$} &
\multicolumn{1}{c}{$N_{Dis}$} & \multicolumn{1}{c}{$T_{Insp}$} & $\Delta T_{Insp}$ & \multicolumn{1}{c}{$\Ms$} & \multicolumn{1}{c}{$\Rs$} & 
 \multicolumn{1}{c}{$R_t$} & \multicolumn{1}{c}{$r_{pi}$} & $r_{pi}/R_t$ & 
\multicolumn{1}{c}{a} & \multicolumn{1}{c}{$e$} & \multicolumn{1}{c}{$r_{pf}$}\\ 
 & rel. & [$M_\odot$] & [$M_\odot$] & & \multicolumn{1}{c}{[Myr]} & [yrs] & \multicolumn{1}{c}{[$M_\odot$]} & \multicolumn{1}{c}{[$R_\odot$]} & 
 \multicolumn{1}{c}{[$R_\odot$]} & 
 [$R_\odot$] &  & \multicolumn{1}{c}{[$R_\odot$]} &  & \multicolumn{1}{c}{[$R_\odot$]}\\
\noalign{\smallskip}
 1   & IMK99 & 3572.4 & 3667.5 & 29 & 3.16 & $1.1\cdot10^2$ &  6.12 &  3.0 &  26.2 &  26.5 & 1.01 &  13375 & 0.998 &  53.0 \\
     &       &        &        &    & 8.94 & $1.2\cdot10^5$ &  1.37 &  1.4 &  19.3 &  58.4 & 3.03 &   1408 & 0.959 &  90.2 \\
     &       &        &        &    &10.12 & $1.7\cdot10^3$ &  2.02 &  1.6 &  20.4 &  21.8 & 1.07 &  34149 & 0.999 &  58.9 \\
 2   & IMK99 & 4076.3 & 4188.9 & 27 & 3.05 & $2.5\cdot10^2$ &  1.49 &  1.4 &  20.8 &  35.1 & 1.69 &  33410 & 0.999 &  42.9 \\ 
     &       &        &        &    & 8.92 & $1.9\cdot10^2$ &  1.77 &  1.5 &  21.1 &  36.6 & 1.73 & 166379 & 0.999 &  38.8 \\
 3   & IMK99 & 2786.2 & 3232.7 & 31 & 3.13 & $5.7\cdot10^3$ & 64.97 & 44.4 & 161.1 & 383.7 & 2.38 & 14695  & 0.974 & 728.2 \\  
     &       &        &        &    & 3.41 & $7.5\cdot10^3$ & 50.42 & 30.5 & 121.6 & 169.8 & 1.40 & 55688  & 0.997 & 336.4 \\
     &       &        &        &    & 4.80 & $6.7\cdot10^3$ & 25.61 &1961.8&10082.9&16172.1& 1.60 &315615  & 0.949 &20475.5\\
     &       &        &        &    & 6.13 & $1.5\cdot10^3$ & 17.64 &  8.3 &  48.3 & 113.7 & 2.36 & 22883  & 0.995 & 120.3 \\
     &       &        &        &    &10.03 & $2.6\cdot10^5$ &  1.35 &  1.3 &  18.3 &  36.2 & 1.98 & 76953  & 0.999 &  69.2 \\
 4   & IMK99 & 3325.3 & 3783.8 & 40 & 3.13 & $3.7\cdot10^2$ & 48.84 & 22.5 &  96.2 & 149.1 & 1.55 &100084  & 0.998 & 196.9 \\ 
     &       &        &        &    & 3.42 & $2.0\cdot10^4$ & 49.75 & 28.9 & 124.4 & 185.5 & 1.49 & 31597  & 0.994 & 370.9 \\
     &       &        &        &    & 3.77 & $4.6\cdot10^3$ & 39.68 & 20.5 &  95.8 & 130.2 & 1.36 & 21103  & 0.994 & 252.4 \\
     &       &        &        &    &10.91 & $2.0\cdot10^2$ &  3.01 &  2.0 &  22.4 &  23.8 & 1.06 & 25552  & 0.999 &  47.4 \\
 5   & IMK99 & 3190.3 & 3464.0 & 18 & 3.14 & $2.0\cdot10^2$ & 61.72 & 37.5 & 144.7 & 485.1 & 3.35 & 20418  & 0.976 & 447.8 \\
     &       &        &        &    & 3.55 & $3.3\cdot10^3$ &  1.43 &  1.4 &  19.2 &  24.6 & 1.28 &117399  & 0.999 &  51.0 \\
     &       &        &        &    & 5.28 & $6.3\cdot10^1$ &  2.32 &  1.7 &  20.3 &  23.1 & 1.14 & 43542  & 0.999 &  42.9 \\
     &       &        &        &    & 7.74 & $2.7\cdot10^4$ & 21.60 &988.0 &5499.1 &10847.7& 1.97 &372902  & 0.971 &14373.1\\
     &       &        &        &    &11.69 & $3.6\cdot10^4$ & 15.69 &682.7 &4249.3 & 5530.9& 1.30 &363884  & 0.998 & 8503.8\\
 6   & IMK99 & 3603.7 & 4230.2 & 39 & 3.15 & $7.9\cdot10^4$ & 65.53 & 47.3 & 185.5 &  688.0& 3.71 & 9642   & 0.929 &  623.5\\ 
     &       &        &        &    & 3.53 & $3.0\cdot10^3$ & 46.70 & 27.5 & 122.1 &  565.4& 4.63 &215807  & 0.997 &  248.5\\
     &       &        &        &    & 3.59 & $2.9\cdot10^3$ & 29.15 & 11.8 &  61.8 &  154.7& 2.50 &133332  & 0.999 &  199.1\\
     &       &        &        &    & 4.45 & $4.9\cdot10^3$ & 37.05 & 31.2 & 152.1 &  192.8& 1.27 & 82735  & 0.998 &  385.4\\
     &       &        &        &    & 4.98 & $2.5\cdot10^3$ & 25.41 & 12.7 &  70.9 &  120.8& 1.70 & 49705  & 0.998 &  222.3\\
     &       &        &        &    & 8.01 & $2.0\cdot10^3$ &  9.86 &1489.9&11479.4&14360.4& 1.25 &655428  & 0.978 &22904.6\\
\noalign{\smallskip}
 7   & BAC84 & 1980.7 & 2597.8 & 28 & 3.04 & $1.5\cdot 10^4$& 48.36 & 25.1 &  89.1 & 443.2 & 4.97 & 19506  & 0.977 & 222.2 \\ 
     &       &        &        &    & 3.84 & $6.8\cdot 10^4$& 40.74 & 42.5 & 166.0 & 348.5 & 2.10 & 30596  & 0.989 & 505.2 \\ 
     &       &        &        &    & 5.79 & $6.6\cdot 10^3$&  2.05 &  1.6 &  17.7 &  25.2 & 1.42 & 45765  & 0.999 &  51.7 \\
 8   & BAC84 & 2064.6 & 2383.1 & 19 & 3.04 & $3.2\cdot 10^2$& 62.04 & 34.9 & 116.0 & 126.3 & 1.09 &157896  & 0.999 & 248.2 \\ 
     &       &        &        &    & 3.40 & $1.2\cdot 10^3$& 43.49 & 20.6 &  79.7 & 385.2 & 4.83 & 56733  & 0.993 & 168.7 \\
 9   & BAC84 & 1147.2 & 1579.6 & 19 & 3.42 & $2.2\cdot 10^4$& 44.00 & 27.2 &  85.2 & 191.4 & 2.25 & 21077  & 0.991 & 265.7 \\
     &       &        &        &    & 4.11 & $3.2\cdot 10^2$& 37.93 & 32.8 & 112.7 & 193.8 & 1.72 &   251  & 0.229 & 238.1 \\ 
     &       &        &        &    & 4.95 & $7.4\cdot 10^3$& 27.91 & 17.3 &  67.7 & 177.1 & 2.62 & 27882  & 0.994 & 195.4 \\
     &       &        &        &    & 6.38 & $4.4\cdot 10^5$& 12.45 &  5.5 &  28.5 &  93.0 & 3.27 & 33949  & 0.997 & 130.2 \\
     &       &        &        &    &11.82 & $1.8\cdot 10^5$& 12.81 &843.1 &4328.9 &11232.9& 2.59 &294365  & 0.962 &12043.5\\ 
10   & BAC84 & 1655.3 & 2009.4 & 31 & 3.16 & $1.1\cdot 10^4$& 50.05 & 55.3 & 185.0 & 250.7 & 1.35 & 70838  & 0.996 & 368.1 \\
     &       &        &        &    & 4.42 & $1.8\cdot 10^3$& 35.02 & 31.9 & 121.3 & 239.7 & 1.98 & 22510  & 0.989 & 264.7 \\
     &       &        &        &    & 4.93 & $7.9\cdot 10^3$& 31.89 & 32.4 & 128.2 & 293.7 & 2.29 &  3538  & 0.917 & 284.5 \\
     &       &        &        &    & 5.12 & $6.2\cdot 10^3$& 28.84 &1738.4&7161.1 &22012.6& 3.07 &190684  & 0.885 &14997.0\\
     &       &        &        &    & 5.83 & $1.5\cdot 10^5$& 24.35 & 15.6 &  68.5 & 195.8 & 2.86 & 45854  & 0.996 & 255.2 \\
     &       &        &        &    & 6.33 & $2.6\cdot 10^4$& 13.34 &  5.9 &  31.9 &  68.3 & 2.14 &145350  & 0.999 & 105.7 \\
     &       &        &        &    & 9.90 & $8.4\cdot 10^4$& 16.94 & 15.1 &  75.8 & 179.5 & 2.37 & 44328  & 0.996 & 256.5 \\
     &       &        &        &    &11.13 & $1.4\cdot 10^5$& 16.13 &708.0 &3629.3 &5552.3 & 1.53 & 80890  & 0.931 &9856.8 \\
11   & BAC84 & 1207.3 & 1565.3 & 19 & 5.36 & $7.4\cdot 10^5$& 30.87 & 27.8 & 104.3 & 189.4 & 1.82 & 49832  & 0.996 & 417.8 \\
     &       &        &        &    & 6.00 & $6.8\cdot 10^4$& 24.54 &1357.5&5524.5 &22582.5& 4.09 &104937  & 0.785 &16732.5\\
     &       &        &        &    &10.06 & $5.1\cdot 10^2$& 16.64 & 14.1 &  65.9 &  81.3 & 1.23 & 41275  & 0.998 & 139.7 \\
12   & BAC84 & 1700.8 & 2287.7 & 20 & 3.30 & $3.2\cdot 10^2$& 27.27 & 10.5 &  44.2 &  92.4 & 2.09 &  5641  & 0.984 &  90.0 \\
     &       &        &        &    & 8.32 & $1.2\cdot 10^4$& 11.89 &1385.5&8154.8 &9569.0 & 1.17 &212502  & 0.955 &16101.8\\
     &       &        &        &    &10.12 & $2.5\cdot 10^4$& 16.99 &  58.9& 309.9 & 924.1 & 2.98 & 49761  & 0.981 & 1629.2\\
     &       &        &        &    &11.37 & $1.1\cdot 10^3$&  2.33 &   1.7&  17.9 &  43.7 & 2.44 & 89189  & 0.999 &   44.2\\
\end{tabular}
\end{table*}

%% file: paper.bbl
\begin{thebibliography}{}

\bibitem[ 1999]{AA99}Aarseth, S. J., 1999, PASP, 111, 1333
\bibitem[ 1999]{A99}Alexander, T., 1999, ApJ, 520, 137 
\bibitem[ 2001]{AL01}Alexander \& Livio, 2001, ApJ, 560, L143
\bibitem[ 2003]{AH03}Alexander, T., Hopman, C., 2003, ApJ, 590, L29
\bibitem[ 2003]{AM03}Alexander, T., Morris, M., 2003, ApJ, 590, L25
\bibitem[ 1976]{BW76}Bahcall, J. N., Wolf, R. A., 1976, ApJ, 209, 214
\bibitem[ 1977]{BW77}Bahcall, J. N., Wolf, R. A., 1977, ApJ, 216, 883
\bibitem[ 2004a]{Baum04a}Baumgardt, H., Makino, J., Ebisuzaki, T., 2004a, ApJ, 613, 1133 
\bibitem[ 2004b]{Baum04b}Baumgardt, H., Makino, J., Ebisuzaki, T., 2004b, ApJ, 613, 1143 
\bibitem[ 2005]{B05}Blecha, L., et al., 2005, ApJ submitted, astro-ph/0508597
\bibitem[ 1984]{BO84}Bond, J. R., Arnett, W. D., Carr, B. J., 1984, ApJ, 280, 825
\bibitem[ 1978]{CK78}Cohn, H., Kulsrud, R. M. 1978, ApJ, 226, 1087
\bibitem[ 2005]{Cea05}Cuadra, J., Nayakshin, S., Springel, V., Di Matteo, T., 2005, MNRAS in press, astro-ph/0502044
\bibitem[ 2004]{FT04}Fiorito, R., Titarchuk, L., 2004, ApJ, 614, L113
\bibitem[ 1976]{FR76}Frank, J., Rees, M. J., 1976, MNRAS, 176, 633
\bibitem[ 2001a]{FR01a}Freitag, M., 2001, A\&A, 375, 711
\bibitem[ 2001b]{FR01b}Freitag, M., 2001, Class. Quantum Grav., 18, 4033
\bibitem[ 2003]{FR03}Freitag, M., 2003, ApJ, 583, L21
\bibitem[ 2005]{FR05}Freitag, M., G\"urkan, M. A., Rasio, F. A., 2005, MNRAS submitted, astro-ph/0503130
\bibitem[ 2005]{GU05}Guti\'errez, C. M., L\'opez-Corredoira, M., 2005, ApJ, 622, 89
\bibitem[ 1995]{HB95}Hils, D., Bender, P. L., 1995, ApJ, 445, L7
\bibitem[ 2004]{HPZA04}Hopman, C., Portegies Zwart, S.F., Alexander, T., 2004, ApJ, 604, L101
\bibitem[ 2005]{HA05}Hopman, C., Alexander, T., 2005, ApJ, 629, 362
\bibitem[ 2005]{HPZ05}Hopman, C., \& Portegies Zwart, S. F., 2005, MNRAS,
363, L56
\bibitem[ 2000]{HU00}Hurley J.\ R., Pols O.\ R., Tout C.\ A., 2000, MNRAS 315, 543
\bibitem[ 2004]{IW04}Irwin, J. A., Bregman, J. N., \& Athey, A. E., 2004, ApJ, 601, L143
\bibitem[ 1999]{IS99}Ishii, M., Ueno, M., Kato, M., 1999, PASJ, 51, 417
\bibitem[ 2002]{IV02}Ivanov, P. B., 2002, MNRAS, 336, 373, 2002
\bibitem[ 2001]{KA01}Kaaret, P., et al., 2001, MNRAS 321, L29 

\bibitem[ 2001]{KI01}King A. R., Davies, M. B., Ward, M. J., Fabbiano, G., 
  Elvis, M., 2001, ApJ, 552, 109
\bibitem[ 2002]{KFM02}K\"ording, E., Falcke, H., Markoff, S., 2002, A\&A, 382, L13  
\bibitem[ 1992]{KO92}Kochanek, C. S., 1992, ApJ, 385, 604  
\bibitem[ 2005]{K05}Kuntz, K. D., Gruendl, R. A., Chu, Y-H, Chen, C.-H., Still, M., Mukai, K., \& Mushotzky, R. F., 2005, ApJ, 620, L31
\bibitem[ 1965]{KU65} Kustaanheimo, P., Stiefel, E. L., 1965, J.\ Reine Angew.\ Math., 218, 204
\bibitem[ 1986]{LE86}Lee, H. M., Ostriker, J. P., 1986, ApJ, 310, 176
\bibitem[ 2004]{L04}Li, Xiang-Dong, ApJ, 2004, 616, L119
\bibitem[ 1977]{LS77}Lightman, A. P., Shapiro, S. L., 1977, ApJ, 211, 244
\bibitem[ 2004]{LBS04}Liu, J-F, Bregman, J. N., \& Seitzer, P., 2004, ApJ, 602, L249
\bibitem[ 2005]{M05}Madhusudhan, N., et al., 2005, ApJ submitted, astro-ph/0511393 
\bibitem[ 2003]{MA03}Makino, J., Fukushige, T., Koga, M., \& Namura, K., 2003, PASJ, 55, 1163 
\bibitem[ 1987]{McM87}McMillan, S. L. W., McDermott, P. N., Taam,
R. E. 1987, ApJ, 318, 261
\bibitem[ 1999]{MT99}Magorrian, J., Tremaine, S., 1999, MNRAS,309, 447
\bibitem[ 2001]{MH01} Miller, M. C. \& Hamilton, D. P., 2001, MNRAS, 330, 232
\bibitem[ 2003]{MC03}McCrady, N., Gilbert, A. M., Graham, J. R., 2003, ApJ, 596, 240
\bibitem[ 2004]{MFM04}Miller, J. M., Fabian, A. C., Miller, M. C., 2004, ApJ 614, L117 
\bibitem[ 2004]{MC04} Miller, M. C., \& Colbert, E. J. M., 2004,  International Journal of Modern Physics D., 13, 01
\bibitem[ 2000]{MG00}Miralda-Escud\'{e}, J. \& Gould, A., 2000, ApJ,
545, 847
\bibitem[ 2001]{MR01}Madau, P., \& Rees, M. J., 2001, ApJ, 551, L27
\bibitem[ 2001]{MA01}Matsumoto, H., et al., 2001, ApJ, 547, L25
\bibitem[ 2004]{MO04}Mouawad, N., Eckart, A., Pfalzner, S., Sch\"{o}del, R., Moultaka, J., Spurzem, R., 2004, Astronomische Nachrichten Vol. 326, 2, 83-95
\bibitem[ 1964]{PE64}Peters, P. C., 1964, Physical Reviews B, 136, 1224
\bibitem[ 1996]{P96} Podsiadlowski, Ph., 1996, MNRAS, 279, 1104
\bibitem[ 1999]{PZea99}Portegies Zwart, S. F., Makino, J., McMillan, S. L. W., \& Hut, P., 1999, A\&A, 348, 117
\bibitem[ 2002]{PZMcM02}Portegies Zwart, S. F., McMillan, S. L. W., 2002 {\bf fillin}
\bibitem[ 2004]{PZea04}Portegies Zwart, S. F., Baumgardt, H., Makino, J., McMillan, S. L., Hut, P., 2004, Nature, 428, 724
\bibitem[ 2005]{PZea05}Portegies Zwart, S. F., Baumgardt, H., McMillan, S. L., Makino, J., Hut, P., Ebisuzaki, t., 
  2005, ApJ submitted, astro-ph/0511397
\bibitem[ 2004]{PZDM04}Portegies Zwart, S. F., Dewi, J., \& Maccarone, T., 2004, MNRAS, 355, 413
\bibitem[ 1993]{PZ93}Portegies Zwart, S. F., Meinen, A. T., 1993, A\&A, 280, 174
\bibitem[ 1977]{PT77}Press, W. H., Teukolsky, S. A., ApJ, 213, 183
\bibitem[ 2004]{P04}Preto, M., Merritt, D., Spurzem, R., 2004, ApJ, 613, 109  
\bibitem[ 2005]{RA05}Rappaport, S. A., Podsiadlowski, P., Pfahl, E., 2005, MNRAS, 356, 401
\bibitem[ 1987]{RA87}Ray, A., Kembhavi, A. K., Antia, H. M., 1987, A\&A, 184, 164
\bibitem[ 1988]{R88} Rees, M. J., 1988, Nature, 333, 523
\bibitem[ 2001]{SO04} Soria, R., Cropper, M., Motch, C., 2004, Chinese Journal 
  of Ast.\ \& Astrophys.\ in press, astro-ph/040913
\bibitem[ 1997]{SR97}Sigurdsson, S., Rees, M. J., 1997, MNRAS 284, 318
\bibitem[ 2003]{SM03}Stohmayer, T. E., Mushotzky, R. F., 2003, ApJ, 586, 61
\bibitem[ 2004]{S04} Swartz, D. A., Ghosh, K. K., Tennant, A. F., \& Wu, K., 2004, ApJ, 154, S519

\bibitem[ 1999]{SU99}Syer, D., Ulmer, A., 1999, MNRAS, 306, 35
\bibitem[ 2004]{WM04}Wang, J., Merritt, D., 2004, ApJ, 600, 149

\end{thebibliography}
